\documentclass[journal,10pt]{IEEEtran}
\usepackage{graphicx}
\usepackage{caption,color}
\usepackage{subcaption}
\usepackage{refstyle,cite}
\usepackage{amsthm,amssymb}
\usepackage{amsmath}
\usepackage{algorithm}
\usepackage{algorithmic}
\usepackage{epstopdf}
\usepackage{cuted}

\ifCLASSINFOpdf
\else
\fi
\begin{document}
\linespread{1.02}
\title{Study of Relay Selection  for Physical-Layer Security in Buffer-Aided Relay Networks Based on the Secrecy Rate Criterion}

\author{Xiaotao Lu and Rodrigo C. de~Lamare
\thanks{Xiaotao Lu is with the Communications Research Group, Department of Electronics, University of York, YO10 5DD York, U.K., R. C. de Lamare is with CETUC, PUC-Rio, Brazil and with the
Communications Research Group, Department of Electronics, University
of York, YO10 5DD York, U.K. e-mails: xtl503@york.ac.uk;
rodrigo.delamare@york.ac.uk.}}

\maketitle
\begin{abstract}
In this paper, we investigate an opportunistic relay and jammer
scheme along with relay selection algorithms based on the secrecy
rate criterion in multiple-input multiple-output buffer-aided down
link relay networks, which consist of one source, a number of relay
nodes, legitimate users and eavesdroppers, with the constraints of
physical layer security. The opportunistic relay and jammer scheme
is employed to improve the transmission rate and different relay
selection policies are performed to achieve better secrecy rate with
the consideration of eavesdroppers. Among all the investigated relay
selection policies, a relay selection policy which is developed to
maximize the secrecy rate based on exhaustive searches outperforms
other relay selection policies in terms of secrecy rate. Based on
the secrecy rate criterion, we develop a relay selection algorithm
without knowledge of the channels of the eavesdroppers. We also
devise a greedy search algorithm based on the secrecy rate criterion
to reduce the computational complexity of the exhaustive search
technique. Simulations show the superiority of the secrecy rate
criterion over competing approaches.

\end{abstract}


\section{Introduction}

Secure transmission is difficult to achieve in broadcast channels
due to the nature of wireless communications. Traditional encryption
techniques are implemented in the network layer with complex
algorithms and high cost. To reduce such cost, researchers are
exploring novel security techniques in the physical layer.
Physical-layer security has been first illustrated by Shannon using
an information theoretic viewpoint \cite{Shannon}. The feasibility
of physical-layer security has been discussed by Shannon in a
theoretical level in \cite{Shannon}. Later on in \cite{Wyner} a
wire-tap channel, which can achieve positive secrecy rate, has been
proposed by Wyner under the assumption that the users have a better
statistical channel than eavesdroppers. Since then further research
has been devoted to the wire-tap model and techniques such as
broadcast channels \cite{Csiszar}, MIMO channels, artificial noise,
beamforming as well as relay techniques. This paper focuses on relay
techniques.

Recently, the concept of physical-layer security with multiuser
wireless networks has been investigated \cite{Mukherjee}. Relay
systems are an important evolution of secure transmission strategies
and techniques to further improve the performance of relay systems
such as relay selection \cite{tds} and buffer-aided relay nodes
\cite{Zlatanov} are drawing significant attention. Opportunistic
relay schemes have been applied to buffer-aided systems
\cite{Nomikos1}, \cite{Nomikos2} and \cite{Lee}. In opportunistic
relay schemes, the inter-relay interference (IRI) is an important
aspect that should be taken into account.

In our previous work \cite{Xiaotao3}, we have introduced an
opportunistic relay and jammer scheme and investigated its potential
for improving secrecy rate. In this work, we employ the same scheme
and focus our research on different relay selection algorithms
\cite{Jing,Clarke,Ding,Song,Talwar}. Unlike prior art which relies
on the signal-to-interference-plus-noise (SINR) and channel state
information \cite{Jingchao} approaches, we examine the potential of
using the secrecy rate as the criterion for the selection of relays.
In particular, a relay selection strategy is developed to maximize
the secrecy rate based on exhaustive searches. A greedy search
algorithm is then developed to reduce the computational complexity
of the exhaustive search approach.

This paper is organized as follows. Section II details the system
model, describes the opportunistic relay and jammer buffer-aided
relay system. We focus on the transmission from the source to the
relays and from the relays to the users. Section III presents all
investigated conventional as well as the proposed relay selection
criterion, whereas Section IV details the proposed secrecy rate
based criterion with partial channel information. Section V shows
and discusses the numerical results, while the conclusions are drawn
in Section VI.

\section{System Model and Performance Metrics}

In this section, a brief introduction of the buffer-aided relay system model is
given to describe the data transmission. The performance metrics illustrate the
assessment of the proposed and existing techniques described in this paper.

\subsection{System Model}

\begin{center}
\begin{figure}[h]
\centering
\includegraphics[scale=0.75]{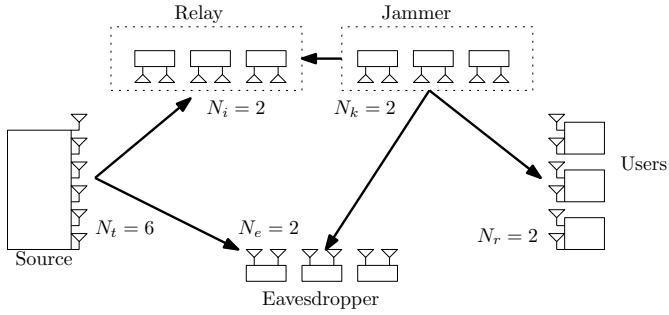}
\vspace{0.5em} \caption{System model of a MU-MIMO system with $M$
users, $N$ eavesdroppers $T$ relays and $K$ jammers} \label{fig:sys}
\end{figure}
\end{center}

Fig. \ref{fig:sys} gives a description of a source node with $N_{t}$
antennas that transmits the data streams to $M$ users in the
presence of $N$ eavesdroppers. With $T$ relays and $K$ jammers, in
each time slot, the selected relays will receive signals from both
the source and the jammers. Each relay and jammer is equipped with
$N_{i}$ and $N_{k}$ antennas. At the receiver side each user and
eavesdropper is equipped with $N_{r}$ and $N_{e}$ receive antennas.
The quantities ${\boldsymbol H}_{i}\in {\mathbb{C}}^{N_{i}\times
N_{t}}$ and ${\boldsymbol H}_{e}\in {\mathbb{C}}^{N_{e}\times
N_{t}}$ denote the channel matrices of the $i$th relay and the $e$th
eavesdropper, respectively. The quantities ${\boldsymbol H}_{ke}\in
{\mathbb{C}}^{N_{e}\times N_{k}}$ and ${\boldsymbol H}_{kr}\in
{\mathbb{C}}^{N_{r}\times N_{k}}$ denote the channel matrices of the
$e$th eavesdropper to the $k$th jammer and the $r$th user to the
$k$th jammer, respectively. The channel between the $k$th relay to
the $i$th relay is modeled by ${\boldsymbol H}_{ki}\in
{\mathbb{C}}^{N_{i}\times N_{k}}$. To support $M$ users
transmission, the source is equipped with $N_{t}\geqslant N_{r}M$
antennas.

The vector ${\boldsymbol s}_{r}^{(t)} \in {\mathbb{C}}^{N_{r}\times
1}$ represents the data symbols to be transmitted corresponding to
each user in time slot $t$. The total transmit signal at the
transmitter can be expressed as
\begin{equation}{\boldsymbol
s}^{(t)}={\left[ {{\boldsymbol s}_{1}^{(t)}}^{T} \quad {{\boldsymbol
s}_{2}^{(t)}}^{T} \quad {{\boldsymbol s}_{3}^{(t)}}^{T} \quad \cdots
\quad {{\boldsymbol s}_M^{(t)}}^{T}\right]}^{T}.
\end{equation}
In prior work, several precoding techniques have been considered to
eliminate the interference between different users
\cite{Gaojie,Xiaotao1,Keke1,Keke2,Keke3,sint}. Alternative
approaches to the computation of the transmit filters can be
employed
\cite{int,Chen,Meng,l1cg,zhaocheng,alt,jiolms,jiols,jiomimo,jidf,fa10,saabf,barc,honig,mswfccm,song,locsme}..
In this work, we use linear zero-forcing precoding and the precoding
matrix is described by
\begin{equation}
\boldsymbol U_{i}={\boldsymbol H_{i}}^{H}({\boldsymbol
H_{i}}{\boldsymbol H_{i}}^{H})^{-1}.
\end{equation}

The channels of the selected jammers to the $r$th user are given by
\begin{equation}
{\boldsymbol H_{r}}^{(t)}={\left[ {{\boldsymbol
H}_{k_{1}r}^{(t)}}^{T} \quad {{\boldsymbol H}_{k_{2}r}^{(t)}}^{T}
\quad {{\boldsymbol H}_{k_{3}r}^{(t)}}^{T} \quad \cdots \quad
{{\boldsymbol H}_{k_{M}r}^{(t)}}^{T}\right]}.
\end{equation}
The channels of the jammers to the $i$th relay channel are
\begin{equation}
{\boldsymbol H_{Ki}}^{(t)}={\left[ {{\boldsymbol
H}_{k_{1}i}^{(t)}}^{T} \quad {{\boldsymbol H}_{k_{2}i}^{(t)}}^{T}
\quad {{\boldsymbol H}_{k_{3}i}^{(t)}}^{T} \quad \cdots \quad
{{\boldsymbol H}_{k_{M}i}^{(t)}}^{T}\right]}.
\end{equation}
To simplify the calculation, we assume that each relay will have the
same antenna as each user which means ${\boldsymbol
s}_{i}^{(t)}={\boldsymbol s}_{r}^{(t)}$. In each phase, the received
signal ${\boldsymbol y}_{i}^{(t)}\in {\mathbb{C}}^{N_{i}\times 1}$
at each relay node can be expressed as:
\begin{equation}
{\boldsymbol y}_{i}^{(t)} = {\boldsymbol H}_{i}\boldsymbol U_{i}{\boldsymbol
s}_{i}^{(t)}+\sum_{j\neq i}{\boldsymbol H}_{i}\boldsymbol U_{j}{\boldsymbol
s}_{j}^{(t)}+{\boldsymbol H}_{Ki}^{(t)}{\boldsymbol
y}_{k}^{(pt)}+\boldsymbol{n}_{i} \label{eqn:yit}
\end{equation}
In (\ref{eqn:yit}), the value $pt$ represents the previous time slot when the
signal is stored as a jamming signal in the buffer at the relay nodes. The term
${\boldsymbol H}_{Ki}{\boldsymbol y}_{k}^{(pt)}$ is regarded as the inter-relay
interference (IRI) between the $i$th relay and $K$ jammers and ${\boldsymbol
y}_{k}^{(pt)}$ is determined as the jamming signal according to a SINR
criterion as in \cite{Nomikos1}. With the theorem in \cite{Nomikos1}, the IRI
can be eliminated.

The channel of the jammers to the $e$th eavesdropper is described by
\begin{equation}
{\boldsymbol H_{Ke}}^{(t)}={\left[ {{\boldsymbol
H}_{k_{1}e}^{(t)}}^{T} \quad {{\boldsymbol H}_{k_{2}e}^{(t)}}^{T}
\quad {{\boldsymbol H}_{k_{3}e}^{(t)}}^{T} \quad \cdots \quad
{{\boldsymbol H}_{k_{M}e}^{(t)}}^{T}\right]}.
\end{equation}
The received signal at the eth eavesdropper is given by
\begin{equation}
{\boldsymbol y}_{e}^{(t)}={\boldsymbol H}_{e}\boldsymbol U_{i}{\boldsymbol
s}_{i}^{(t)} +\sum_{j\neq i}{\boldsymbol H}_{e}\boldsymbol U_{j}{\boldsymbol
s}_{j}^{(t)}+{\boldsymbol H}_{Ke}^{(t)}{\boldsymbol
y}_{k}^{(pt)}+\boldsymbol{n}_{e}. \label{eqn:yet}
\end{equation}
For the eavesdropper, the term ${\boldsymbol H}_{Ke}^{(t)}{\boldsymbol
y}_{k}^{(pt)}$ acts as the jamming signal and this jamming signal can not be
removed without the knowledge of the channel from the kth jammer to the eth
eavesdropper.

In (\ref{eqn:yit}) and (\ref{eqn:yet}), the IRI term between the
relay nodes or the jamming signal to the eavesdropper is
simultaneously the transmit signal from the relays nodes to the
destination. We assume that the transmit signal from the relay nodes
is given by
\begin{equation}
{\boldsymbol r}^{(t)}={\left[ {{\boldsymbol y}_{1}^{(pt_{1})}}^{T}
\quad {{\boldsymbol y}_{2}^{(pt_{2})}}^{T} \quad {{\boldsymbol
y}_{3}^{(pt_{3})}}^{T} \quad \cdots \quad {{\boldsymbol
y}_{T}^{(pt_{T})}}^{T}\right]}^{T}.
\end{equation}
Note that $pt$ represents the previous time slot and due to the
characteristics of buffer relay nodes, the values can be different
for each relay node. The received signal at the destination can be
expressed as:
\begin{equation}
{\boldsymbol y}_{r}^{(t)}={\boldsymbol H}_{r} {\boldsymbol
y}_{k}^{(pt_{k})}+\sum_{j\neq r}{\boldsymbol H}_{r}{\boldsymbol
y}_{j}^{(pt_{j})}+\boldsymbol{n}_{r}. \label{eqn:yrt}
\end{equation}
Advanced detection techniques can be used to improve the quality of
the reception \cite{spa,mfsic,jiomimo,mbdf}

\section{Selection with Jamming Function Relays in Multiuser MIMO Buffer-aided Relay System}

In this section, a novel selection approach with jamming function relays is
introduced. We first consider a simple single-antenna scenario and then the
selection approach is extended to a MIMO scenario. After that, a further
exploration of the relay selection in a multiuser MIMO buffer-aided relay
system is undertaken.

\subsection{Relay Selection Criteria}

\subsubsection{Conventional Relay Selection Criterion}

The conventional selection relies only on the knowledge of channel
information between the source to the relays and the relays to the
users. In \cite{Krikidis}, a max-min relay selection is considered
as the optimal selection scheme for conventional decode-and-forward
(DF) relay setups. With a single-antenna scenario the relay
selection policy is given as:
\begin{equation}
R_{i}^{*}=\rm{arg} \max_{R_{i} \in \boldsymbol{\Psi}}
\min(\|h_{S,R_{i}}\|^{2},\|h_{R_{i},D}\|^{2}), \label{eqn:maxm}
\end{equation}
where $h_{S,R_{k}}$ is the link between the source to the relay and
$h_{R_{k},D}$ is the relay to the destination.

With the consideration of the eavesdropper, a max-ratio selection policy is
proposed in \cite{Gaojie} and given by
\begin{equation}
R_{i}^{\rm{max-ratio}}=\rm{arg} \max_{R_{i} \in \boldsymbol{\Psi}}\left( \eta_{\rm{max-ratio}_{1}},\eta_{\rm{max-ratio}_{2}}\right)
\label{eqn:maxr}
\end{equation}
with
\begin{equation}
\eta_{\rm{max-ratio}_{1}}=\frac{\max_{R_{i} \in \boldsymbol{\Psi}:\varphi{(Q_{p})}\neq
L}\|h_{S,R_{i}}\|^{2}}{\|h_{se}\|^{2}}
\label{eqn:eta1}
\end{equation}
\begin{equation}
\eta_{\rm{max-ratio}_{2}}=\max_{R_{i} \in \boldsymbol{\Psi}:\varphi{(Q_{p})}\neq
0}\frac{\|h_{R_{i},D}\|^{2}}{\|h_{R_{i}e}\|^{2}}
\label{eqn:eta2}
\end{equation}
The aforementioned relay selection are based on the channel state information.
In \cite{Jingchao} there is also a conventional selection as well as optimal
selection based on the SINR criterion.

\subsubsection{Maximum Likelihood (ML) Relay Selection Criterion}

Based on the conventional relay selection criterion, we substitute the channel
state information with the ML rule which can be expressed as:
\begin{equation}
R_{i}^{\rm{ML}}=\rm{arg} \min_{R_{i} \in \boldsymbol{\Psi}}\left( \eta_{\rm{ML}_{1}},\eta_{\rm{ML}_{2}}\right)
\label{eqn:mlr}
\end{equation}
with
\begin{equation}
\eta_{\rm{ML}_{1}}={\min_{R_{i} \in \boldsymbol{\Psi}:\varphi{(Q_{p})}\neq
L}\|y_{i}-h_{S,R_{i}}s_{i}\|}
\label{eqn:mleta1}
\end{equation}
\begin{equation}
\eta_{\rm{ML}_{2}}=\min_{R_{i} \in \boldsymbol{\Psi}:\varphi{(Q_{p})}\neq
0}{\|y_{r}-h_{R_{i},D}y_{k}^{(pt)}\|}
\label{eqn:mleta2}
\end{equation}
the ML relay selection selects the relay which gives the minimum ML rule value.

\subsubsection{Secrecy Rate Based Relay Selection Criterion}

Similar to the ML relay selection criterion, the secrecy rate (SR)
relay selection criterion is proposed to achieve better secrecy rate
performance. The selection procedure can be expressed as:
\begin{equation}
R_{i}^{\rm{SR}}=\rm{arg} \min_{R_{i} \in \boldsymbol{\Psi}}\left(
\eta_{\rm{SR}_{1}},\eta_{\rm{SR}_{2}}\right) \label{eqn:scr}
\end{equation}
with
\begin{equation}
\eta_{\rm{SR}_{1}}={\max_{R_{i} \in \boldsymbol{\Psi}:\varphi{(Q_{p})}\neq
L}\frac{\|1+h_{S,R_{i}}s_{i}\|}{\|1+h_{se}s_{i}\|}} \label{eqn:sceta1}
\end{equation}
\begin{equation}
\eta_{\rm{SR}_{2}}=\min_{R_{i} \in \boldsymbol{\Psi}:\varphi{(Q_{p})}\neq
0}\frac{\|1+h_{R_{i},D}y_{k}^{(pt)}\|}{\|1+h_{R_{i}e}y_{k}^{(pt)}\|}
\label{eqn:sceta2}
\end{equation}
Based on the single-antenna scenario, the criterion for MIMO system is given by
\begin{equation}
\boldsymbol{\phi}_{m}=\rm{arg} \max_{m \in \boldsymbol{\Psi}} \left(
({\boldsymbol{I}+\boldsymbol{\Gamma}_{e,i}^{(t)}})^{-1}({\boldsymbol{I}+\boldsymbol{\Gamma}_{r,i}^{(t)}})\right),
\label{eqn:sinrct}
\end{equation}
where $\boldsymbol{\Gamma}_{r,i}^{(t)}$ is given as:
\begin{equation}
\boldsymbol{\Gamma}_{r,i}^{(t)} = \sum_{k=1}^{k=K}\frac{P}{N_{k}}{\boldsymbol
H}_{kr} {\boldsymbol H}_{kr}^H({\boldsymbol I}+\frac{P}{N_{t}}{\boldsymbol
H}_{i}^{(pt)} {{\boldsymbol H}_{i}^{(pt)}}^H ). \label{eqn:RR2}
\end{equation}
and
\begin{equation}
\boldsymbol{\Gamma}_{e,i}^{(t)} = ({\boldsymbol I
+\boldsymbol\varDelta})^{-1}{\frac{P}{N_{t}}{\boldsymbol
H}_{e} {{\boldsymbol H}_{e}}^H}, \label{eqn:RE2}
\end{equation}
where
\begin{equation}
\boldsymbol\varDelta = {
\sum_{e=1}^{N}\sum_{k=1}^{K}\frac{P}{N_{k}}{\boldsymbol H}_{ke} {\boldsymbol
H}_{ke}^H({\boldsymbol I}+\frac{P}{N_{t}}{\boldsymbol H}_{i}^{(pt)}
{{\boldsymbol H}_{i}^{(pt)}}^H)}. \label{eqn:RE21}
\end{equation}

\subsubsection{Proposed Secrecy Rate Based Relay Selection Criterion without Knowledge of Eavesdroppers}

Based on (\ref{eqn:yit}) and (\ref{eqn:yet}), the covariance matrix
of the interference and the signal can be obtained as $\boldsymbol
R_{I}=\sum_{j\neq i}\boldsymbol U_{j}{\boldsymbol
s}_{j}^{(t)}{{\boldsymbol s}_{j}^{(t)}}^{H}{\boldsymbol U_{j}}^{H}$
and $\boldsymbol R_{d}=\boldsymbol U_{i}{\boldsymbol
s}_{i}^{(t)}{{\boldsymbol s}_{i}^{(t)}}^{H}{\boldsymbol U_{i}}^{H}$.
When the matrices are square and have equal size we can obtain the
proposed secrecy rate based relay selection criterion as described
by
\begin{align}
\phi_{i}&=\max_{i}
\bigg[\log\big(\det{\left[\boldsymbol I +\boldsymbol R_{I}+(\boldsymbol H_{i}\boldsymbol R_{d}\boldsymbol H_{i}^{H})\right]}\big)\nonumber \\
&\qquad -\log\big(\det{\left[\boldsymbol I +\boldsymbol
R_{d}+(\boldsymbol H_{i}\boldsymbol R_{I}\boldsymbol
H_{i}^{H})\right]}\big)\bigg],  \label{eqn:rnew}
\end{align}
which can be achieved without knowledge of the channels of the
eavesdroppers. The details of the derivation of the above expression
is given in Section IV.

\subsection{Greedy Algorithm in Relay Selection}

When the relay selection criterion is determined, we will give an example using
the greedy search algorithm. Here we choose relays according to the SR
criterion. When the $K$ relays that forward the signals to the users are
selected, the relays used for signal reception are chosen based on the SR
criterion, as given by
\begin{equation}
\boldsymbol{\phi}_{m}=\rm{arg} \max_{m \in \boldsymbol{\Psi}} \left(
({\boldsymbol{I}+\boldsymbol{\Gamma}_{e,i}^{(t)}})^{-1}({\boldsymbol{I}+\boldsymbol{\Gamma}_{m}^{(t)}})\right),
\label{eqn:sinrctgreedy}
\end{equation}
where $\boldsymbol{\phi}_{m}$ represents the selected relays and
$\boldsymbol{\Gamma}_{m}^{(t)}$ is the SINR corresponding to the $m$th relay
which is given by
\begin{equation}
\boldsymbol{\Gamma}_{m}^{(t)}
=({\boldsymbol{I}+\boldsymbol \varDelta_{m}'})^{-1}({\boldsymbol{H}_{m}\boldsymbol{H}_{m}^{H}}),
\label{eqn:gammam}
\end{equation}
where
\begin{equation}
\boldsymbol \varDelta_{m}'=\sum_{k=1}^{K}\boldsymbol{H}_{km}
\boldsymbol{H}_{m}^{(pt)}{\boldsymbol{H}_{m}^{(pt)}}^{H}\boldsymbol{H}_{km}^{H},
\label{eqn:gammam1}
\end{equation}
with the SINR calculated for the $e$th eavesdropper described by
\begin{equation}
\boldsymbol{\Gamma}_{e,i}^{(t)}=({\boldsymbol I
+\boldsymbol \varDelta_{e}'})^{-1}({\frac{P}{N_{t}}{\boldsymbol
H}_{e} {{\boldsymbol H}_{e}}^H}),
\label{eqn:gammae}
\end{equation}
where
\begin{equation}
\boldsymbol
\varDelta_{e}'=\sum_{e=1}^{N}\sum_{k=1}^{K}\frac{P}{N_{k}}{\boldsymbol H}_{ke}
{\boldsymbol H}_{ke}^H({\boldsymbol I}+\boldsymbol{\xi}), \label{eqn:gammae1}
\end{equation}
and
\begin{equation}
\boldsymbol{\xi}=\frac{P}{N_{t}}{\boldsymbol H}_{m}^{(pt)}
{{\boldsymbol H}_{m}^{(pt)}}^H, \label{eqn:xi}
\end{equation}
The main steps are described in Algorithm 1.

\begin{algorithm}
\caption{Greedy Algorithm}
\begin{algorithmic}
\FOR {$k=1:T$}
\FOR {$m=1:M$}
\STATE $\boldsymbol{\Gamma}_{m}^{(t)}
=({\boldsymbol{I}+\boldsymbol \varDelta_{m}'})^{-1}({\boldsymbol{H}_{m}\boldsymbol{H}_{m}^{H}})$
\STATE $\boldsymbol{\xi}=\frac{P}{N_{t}}{\boldsymbol H}_{m}^{(pt)}
{{\boldsymbol H}_{m}^{(pt)}}^H$
\STATE $\boldsymbol{\Gamma}_{e,i}^{(t)}=({\boldsymbol I
+\boldsymbol \varDelta_{e}'})^{-1}({\frac{P}{N_{t}}{\boldsymbol
H}_{e} {{\boldsymbol H}_{e}}^H})$
\STATE $\boldsymbol{\phi}_{m}=\rm{arg} \max_{m \in
\boldsymbol{\Psi}} \det \left(
({\boldsymbol{I}+\boldsymbol{\Gamma}_{e,i}^{(t)}})^{-1}({\boldsymbol{I}+\boldsymbol{\Gamma}_{m}^{(t)}})\right) $
\ENDFOR
\STATE $k=\boldsymbol{\phi}_{m}$
\ENDFOR
\end{algorithmic}
\end{algorithm}

\section{Relay Selection Criterion Analysis}

The details of the conventional relay selection criterion, ML relay
selection criterion as well as the secrecy rate-based relay
selection criterion have been introduced in Section III. Following
this, here we will mainly focus on the proposed secrecy rate based
relay selection criterion with partial channel information. To
simplify the derivation, in the following formulas, we did not take
interference from jammers into consideration and we take the
transmission from the source to the relay as an example. Therefore,
we have
\begin{equation}
\boldsymbol{\Gamma}_{r,i}=(\boldsymbol H_{i}\boldsymbol R_{I}\boldsymbol H_{i}^{H})^{-1}(\boldsymbol H_{i}\boldsymbol R_{d}\boldsymbol H_{i}^{H}),
\end{equation}
and
\begin{equation}
\boldsymbol{\Gamma}_{e,i}=(\boldsymbol H_{e}\boldsymbol R_{I}\boldsymbol H_{e}^{H})^{-1}(\boldsymbol H_{e}\boldsymbol R_{d}\boldsymbol H_{e}^{H}),
\end{equation}
From the original expression for the MIMO secrecy rate performance
and our system structure, we can write the expression as
\begin{equation}
\boldsymbol{\phi}_{i}=\rm{arg} \max_{i \in \boldsymbol{\Psi}}
\left\{
\log[\frac{\det({\boldsymbol{I}+\boldsymbol{\Gamma}_{r,i}})}{\det({\boldsymbol{I}+\boldsymbol{\Gamma}_{e,i}})}]\right\}.
\label{eqn:sinrct2}
\end{equation}
Since the maximization of the $\log$ of an argument is equivalent to
the maximization of the argument (\ref{eqn:sinrct2}) can be
simplified to
\begin{equation}
\boldsymbol{\phi}_{i}=\rm{arg} \max_{i \in \boldsymbol{\Psi}} \left\{
\frac{\det({\boldsymbol{I}+\boldsymbol{\Gamma}_{r,i}})}{\det({\boldsymbol{I}+\boldsymbol{\Gamma}_{e,i}})}\right\},
\label{eqn:sinrct3}
\end{equation}
Similarly to the max-ratio criterion expressed in \cite{Gaojie}, in
order to release the assumption of eavesdroppers channel, instead of
(\ref{eqn:sinrct3}) we consider the following criterion:
\begin{equation}
\boldsymbol{\phi}_{i}=\rm{arg} \max_{i \in \boldsymbol{\Psi}} \left\{
\frac{\det({\boldsymbol{\Gamma}_{r,i}})}{\det({\boldsymbol{\Gamma}_{e,i}})}\right\},
\label{eqn:sinrct4}
\end{equation}
which can be rewritten as
\begin{equation}
\boldsymbol{\phi}_{i}=\rm{arg} \max_{i \in \boldsymbol{\Psi}} \left\{
\frac{\det[(\boldsymbol H_{i}\boldsymbol R_{I}\boldsymbol H_{i}^{H})^{-1}(\boldsymbol H_{i}\boldsymbol R_{d}\boldsymbol H_{i}^{H})]}{\det[(\boldsymbol H_{e}\boldsymbol R_{I}\boldsymbol H_{e}^{H})^{-1}(\boldsymbol H_{e}\boldsymbol R_{d}\boldsymbol H_{e}^{H})]}\right\},
\label{eqn:sinrct41}
\end{equation}
Using linear algebra properties and by further manipulation, we
arrive at
\begin{equation}
\boldsymbol{\phi}_{i}=\rm{arg} \max_{i \in \boldsymbol{\Psi}} \left\{
\frac{\det[(\boldsymbol H_{e}\boldsymbol R_{I}\boldsymbol H_{e}^{H})(\boldsymbol H_{i}\boldsymbol R_{d}\boldsymbol H_{i}^{H})]}{\det[(\boldsymbol H_{i}\boldsymbol R_{I}\boldsymbol H_{i}^{H})(\boldsymbol H_{e}\boldsymbol R_{d}\boldsymbol H_{e}^{H})]}\right\},
\label{eqn:sinrct42}
\end{equation}
If and only if the matrices are square and have equal size, we can
write
\begin{align}
\boldsymbol{\phi}_{i}&=\rm{arg} \max_{i \in \boldsymbol{\Psi}}
\bigg\{\frac{\det(\boldsymbol H_{e})\det(\boldsymbol R_{I})\det(\boldsymbol H_{e}^{H})}{\det(\boldsymbol H_{e})\det(\boldsymbol R_{d})\det(\boldsymbol H_{e}^{H})} \nonumber \\
&\qquad \times \frac{\det(\boldsymbol H_{i}\boldsymbol R_{d}\boldsymbol H_{i}^{H})}{\det(\boldsymbol H_{i}\boldsymbol R_{I}\boldsymbol H_{i}^{H})}\bigg\}
\label{eqn:sinrct43}
\end{align}
In the above expression, the determinant of the channels of the
eavesdroppers can be eliminated, resulting in
\begin{equation}
\boldsymbol{\phi}_{i}=\rm{arg} \max_{i \in \boldsymbol{\Psi}} \left\{
\frac{\det(\boldsymbol R_{I})}{\det(\boldsymbol R_{d})}\frac{\det(\boldsymbol I+\boldsymbol H_{i}\boldsymbol R_{d}\boldsymbol H_{i}^{H})}{\det(\boldsymbol I+\boldsymbol H_{i}\boldsymbol R_{I}\boldsymbol H_{i}^{H})}\right\},
\label{eqn:sinrct44}
\end{equation}
By adding the $\log$ to (\ref{eqn:sinrct44}), we obtain
\begin{equation}
\boldsymbol{\phi}_{i}=\rm{arg} \max_{i \in \boldsymbol{\Psi}} \left\{\log\Big(
\frac{\det(\boldsymbol R_{I})}{\det(\boldsymbol R_{d})}\frac{\det(\boldsymbol I+\boldsymbol H_{i}\boldsymbol R_{d}\boldsymbol H_{i}^{H})}{\det(\boldsymbol I+\boldsymbol H_{i}\boldsymbol R_{I}\boldsymbol H_{i}^{H})}\Big)\right\},
\label{eqn:sinrct45}
\end{equation}
which is equivalent to (\ref{eqn:rnew}). Here we have the results
for equal size channels. When the channels have different sizes we
can add zero elements into the channel matrix and we can still
obtain the same result as equal size channels.

\section{Simulation Results}

In the simulation of the multiuser MIMO scenario, the transmitter is
equipped with $N_{t}=6$ antennas and each relay node is equipped
with $N_{i}=N_{k}=2$ antennas for receiving or transmitting signals.
Each user is equipped with $N_{r}=2$ antennas and the number of
users is set to $M=3$. In this scenario, $N=3$ eavesdroppers
equipped with $N_{e}=2$ antennas each are also considered in the
system. In order to mitigate the interference, a zero-forcing
precoding technique is implemented at the source and also at the
relays. To simplify the transmission scenario, in the selection we
assume the number of selected relays and jammers are the same, which
means we have $T=K$. In all the simulations, all the channels are
generated as flat-fading channel.

\begin{figure}[ht!]
\centering
\includegraphics[width=0.9975\linewidth]{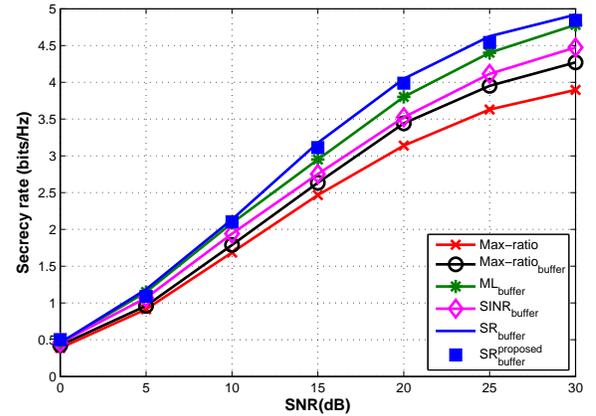}
\vspace{0.5em} \caption{Single-antenna secrecy rate with different
thresholds} \label{fig:thr}
\end{figure}

In Fig. \ref{fig:thr},  different relay selection criteria are
compared in a single-antenna scenario. The secrecy rate performance
has an improvement if buffers are employed in the relay nodes. Among
all the investigated relay selection criteria, SR relay selection
can achieve the best secrecy rate performance. Interestingly, this
approach is typically not used because of the need for knowledge
about the channels of the eavesdroppers.

\begin{figure}[ht!]
\centering
\includegraphics[width=0.9975\linewidth]{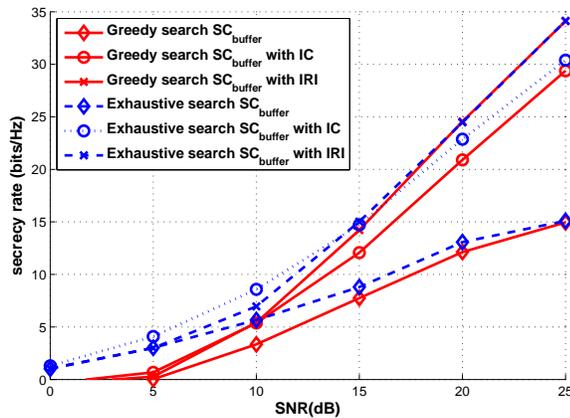}
\vspace{0.5em} \caption{Multi-user system scenario} \label{fig:IRI1}
\end{figure}

In Fig. \ref{fig:IRI1}, with IRI cancellation the secrecy rate is
better than the one without IRI cancellation. Compared with a
single-antenna scenario, the multi-user MIMO system contributes to
the overall improvement in the secrecy rate.

\section{Conclusion}

In this work, we have employed an opportunistic relay and jammer
scheme to enhance the physical-layer secrecy rate performance. The
proposed secrecy rate relay selection policy contributes to the
improvement of the secrecy rate performance. Simulation results
indicate that the secrecy rate criterion relay selection policy
achieves the best secrecy rate performance and the greedy search can
approach a higher secrecy rate performance in a multiuser MIMO relay
system than existing buffer-aided relay systems.

\bibliographystyle{IEEEtran}
\bibliography{referenceIEEE}

\begin{thebibliography}{10}
\providecommand{\url}[1]{#1}
\csname url@samestyle\endcsname
\providecommand{\newblock}{\relax}
\providecommand{\bibinfo}[2]{#2}
\providecommand{\BIBentrySTDinterwordspacing}{\spaceskip=0pt\relax}
\providecommand{\BIBentryALTinterwordstretchfactor}{4}
\providecommand{\BIBentryALTinterwordspacing}{\spaceskip=\fontdimen2\font plus
\BIBentryALTinterwordstretchfactor\fontdimen3\font minus
  \fontdimen4\font\relax}
\providecommand{\BIBforeignlanguage}[2]{{%
\expandafter\ifx\csname l@#1\endcsname\relax
\typeout{** WARNING: IEEEtran.bst: No hyphenation pattern has been}%
\typeout{** loaded for the language `#1'. Using the pattern for}%
\typeout{** the default language instead.}%
\else
\language=\csname l@#1\endcsname
\fi
#2}}
\providecommand{\BIBdecl}{\relax}
\BIBdecl

\bibitem{Shannon}
C.~Shannon, ``Communication theory of secrecy systems,'' \emph{Bell System
  Technical Journal, The}, vol.~28, no.~4, pp. 656--715, Oct 1949.

\bibitem{Wyner}
\BIBentryALTinterwordspacing
A.~D. Wyner, ``The wire-tap channel,'' \emph{Bell System Technical Journal},
  vol.~54, no.~8, pp. 1355--1387, 1975. [Online]. Available:
  \url{http://dx.doi.org/10.1002/j.1538-7305.1975.tb02040.x}
\BIBentrySTDinterwordspacing

\bibitem{Csiszar}
I.~Csiszar and J.~Korner, ``Broadcast channels with confidential messages,''
  \emph{Information Theory, IEEE Transactions on}, vol.~24, no.~3, pp.
  339--348, May 1978.

\bibitem{Mukherjee}
A.~Mukherjee, S.~Fakoorian, J.~Huang, and A.~Swindlehurst, ``Principles of
  physical layer security in multiuser wireless networks: A survey,''
  \emph{Communications Surveys Tutorials, IEEE}, vol.~16, no.~3, pp.
  1550--1573, Third 2014.

\bibitem{tds}
P.~Clarke and R.~C. de~Lamare, ``Transmit diversity and relay selection
  algorithms for multirelay cooperative mimo systems,'' \emph{IEEE Transactions
  on Vehicular Technology}, vol.~61, no.~3, pp. 1084--1098, 2012.

\bibitem{Zlatanov}
N.~Zlatanov and R.~Schober, ``Buffer-aided relaying with adaptive link
  selection-fixed and mixed rate transmission,'' \emph{IEEE Transactions on
  Information Theory}, vol.~59, no.~5, pp. 2816--2840, May 2013.

\bibitem{Nomikos1}
N.~Nomikos, T.~Charalambous, I.~Krikidis, D.~Skoutas, D.~Vouyioukas, and
  M.~Johansson, ``Buffer-aided successive opportunistic relaying with
  inter-relay interference cancellation,'' in \emph{Personal Indoor and Mobile
  Radio Communications (PIMRC), 2013 IEEE 24th International Symposium on},
  Sept 2013, pp. 1316--1320.

\bibitem{Nomikos2}
N.~Nomikos, P.~Makris, D.~Vouyioukas, D.~Skoutas, and C.~Skianis, ``Distributed
  joint relay-pair selection for buffer-aided successive opportunistic
  relaying,'' in \emph{Computer Aided Modeling and Design of Communication
  Links and Networks (CAMAD), 2013 IEEE 18th International Workshop on}, Sept
  2013, pp. 185--189.

\bibitem{Lee}
J.~H. Lee and W.~Choi, ``Multiuser diversity for secrecy communications using
  opportunistic jammer selection: Secure dof and jammer scaling law,''
  \emph{IEEE Transactions on Signal Processing}, vol.~62, no.~4, pp. 828--839,
  Feb 2014.

\bibitem{Xiaotao3}
X.~Lu and R.~C. Lamare, ``Opportunistic relay and jammer cooperation techniques
  for physical-layer security in buffer-aided relay networks,'' in \emph{The
  Twelfth International Symposium on Wireless Communication Systems (ISWCS)},
  2015.

\bibitem{Jing}
Y.~Jing and H.~Jafarkhani, ``Single and multiple relay selection schemes and
  their achievable diversity orders,'' \emph{IEEE Trans. Wireless Commun.},
  vol.~8, no.~3, pp. 1084--1098, Mar 2009.

\bibitem{Clarke}
P.~Clarke and R.~C. de~Lamare, ``Transmit diversity and relay selection
  algorithms for multi-relay cooperative {MIMO} systems,'' \emph{IEEE Trans.
  Veh. Technol}, vol.~61, no.~3, pp. 1084--1098, Mar 2012.

\bibitem{Ding}
M.~Ding, S.~Liu, H.~Luo, and W.~Chen, ``{MMSE} based greedy antenna selection
  scheme for {AF} {MIMO} relay systems,'' \emph{IEEE Signal Process. Lett},
  vol.~17, no.~5, pp. 433--436, May 2010.

\bibitem{Song}
S.~Song and W.~Chen, ``{MMSE} based greedy eigenmode selection for {AF} {MIMO}
  relay channels,'' \emph{IEEE Globecom}, Anaheim, CA, Dec. 2012.

\bibitem{Talwar}
S.~Talwar, Y.~Jing, and S.~Shahbazpanahi, ``Joint relay selection and power
  allocation for two-way relay networks,'' \emph{IEEE Signal Process. Lett},
  vol.~18, no.~2, pp. 91--94, Feb 2011.

\bibitem{Jingchao}
J.~Chen, R.~Zhang, L.~Song, Z.~Han, and B.~Jiao, ``Joint relay and jammer
  selection for secure two-way relay networks,'' \emph{Information Forensics
  and Security, IEEE Transactions on}, vol.~7, no.~1, pp. 310--320, Feb 2012.

\bibitem{Gaojie}
G.~Chen, Z.~Tian, Y.~Gong, Z.~Chen, and J.~Chambers, ``Max-ratio relay
  selection in secure buffer-aided cooperative wireless networks,'' \emph{IEEE
  Transactions on Information Forensics and Security}, vol.~9, no.~4, pp.
  719--729, April 2014.

\bibitem{Xiaotao1}
X.~Lu, K.~Zu, and R.~C.de~Lamare, ``Lattice-reduction aided successive
  optimization tomlinson-harashima precoding strategies for physical-layer
  security in wireless networks,'' in \emph{Sensor Signal Processing for
  Defence (SSPD), 2014}, Sept 2014, pp. 1--5.

\bibitem{Keke1}
K.~Zu and R.~de~Lamare, ``Low-complexity lattice reduction-aided regularized
  block diagonalization for mu-mimo systems,'' \emph{IEEE Communications
  Letters}, vol.~16, no.~6, pp. 925--928, June 2012.

\bibitem{Keke2}
K.~Zu, R.~de~Lamare, and M.~Haardt, ``Generalized design of low-complexity
  block diagonalization type precoding algorithms for multiuser mimo systems,''
  \emph{IEEE Trans. Communications}, vol.~61, no.~10, pp. 4232--4242, October
  2013.

\bibitem{Keke3}
Z.~K, R.~de~Lamare, and M.~Haardt, ``Multi-branch tomlinson-harashima precoding
  design for mu-mimo systems: Theory and algorithms,'' \emph{IEEE Trans.
  Communications}, vol.~62, no.~3, pp. 939--951, March 2014.

\bibitem{sint}
Y.~Cai, R.~C. de~Lamare, and R.~Fa, ``Switched interleaving techniques with
  limited feedback for interference mitigation in ds-cdma systems,'' \emph{IEEE
  Transactions on Communications}, vol.~59, no.~7, pp. 1946--1956, 2011.

\bibitem{int}
R.~C. de~Lamare and R.~C. de~Lamare, ``Adaptive reduced-rank mmse filtering
  with interpolated fir filters and adaptive interpolators,'' \emph{IEEE Signal
  Processing Letters}, vol.~12, no.~3, March 2005.

\bibitem{Chen}
W.~Chen, L.~Dai, K.~B. Letaief, and Z.~Cao, ``A unified cross-layer framework
  for resource allocation in cooperative networks,'' \emph{IEEE Trans. Wireless
  Commun.}, vol.~7, no.~8, pp. 3000--3012, Aug. 2008.

\bibitem{Meng}
R.~Meng, R.~C. de~Lamare, and V.~H. Nascimento, ``Sparsity-aware affine
  projection adaptive algorithms for system identification,'' \emph{in Proc.
  Sensor Signal Processing for Defence Conference}, London, UK, 2011.

\bibitem{l1cg}
Z.~Yang, R.~de~Lamare, and X.~Li, ``Sparsity-aware space-time adaptive
  processing algorithms with l1-norm regularisation for airborne radar,''
  \emph{Signal Processing, IET}, vol.~6, no.~5, pp. 413--423, July 2012.

\bibitem{zhaocheng}
------, ``L1-regularized stap algorithms with a generalized sidelobe canceler
  architecture for airborne radar,'' \emph{Signal Processing, IEEE Transactions
  on}, vol.~60, no.~2, pp. 674--686, Feb 2012.

\bibitem{alt}
R.~C. de~Lamare and R.~C. de~Lamare, ``Sparsity-aware adaptive algorithms based
  on alternating optimization and shrinkage,'' \emph{IEEE Signal Processing
  Letters}, vol.~21, no.~2, pp. 225--229, January 2014.

\bibitem{jiolms}
R.~C. de~Lamare and R.~Sampaio-Neto, ``Reduced--rank adaptive filtering based
  on joint iterative optimization of adaptive filters,'' \emph{IEEE Signal
  Process. Lett.}, vol.~14, no.~12, pp. 980--983, December 2007.

\bibitem{jiols}
------, ``Reduced-rank space--time adaptive interference suppression with joint
  iterative least squares algorithms for spread-spectrum systems,'' \emph{IEEE
  Transactions Vehicular Technology}, vol.~59, no.~3, pp. 1217--1228, March
  2010.

\bibitem{jiomimo}
------, ``Adaptive reduced-rank equalization algorithms based on alternating
  optimization design techniques for {MIMO} systems,'' \emph{IEEE Transactions
  on Vehicular Technology}, vol.~60, no.~6, pp. 2482--2494, July 2011.

\bibitem{jidf}
------, ``Adaptive reduced-rank processing based on joint and iterative
  interpolation, decimation, and filtering,'' \emph{IEEE Transactions on Signal
  Processing}, vol.~57, no.~7, pp. 2503--2514, July 2009.

\bibitem{fa10}
R.~Fa, R.~C. de~Lamare, and L.~Wang, ``Reduced-rank stap schemes for airborne
  radar based on switched joint interpolation, decimation and filtering
  algorithm,'' \emph{IEEE Transactions on Signal Processing}, vol.~58, no.~8,
  pp. 4182--4194, August 2010.

\bibitem{saabf}
S.~Li, R.~C. de~Lamare, and R.~Fa, ``Reduced-rank linear interference
  suppression for ds-uwb systems based on switched approximations of adaptive
  basis functions,'' \emph{IEEE Transactions on Vehicular Technology}, vol.~60,
  no.~2, pp. 485--497, Feb 2011.

\bibitem{barc}
R.~C. de~Lamare, R.~Sampaio-Neto, and M.~Haardt, ``Blind adaptive constrained
  constant-modulus reduced-rank interference suppression algorithms based on
  interpolation and switched decimation,'' \emph{IEEE Transactions on Signal
  Processing}, vol.~59, no.~2, pp. 681--695, Feb 2011.

\bibitem{honig}
M.~L. Honig and J.~S. Goldstein, ``Adaptive reduced-rank interference
  suppression based on the multistage wiener filter,'' \emph{IEEE Transactions
  on Communications}, vol.~50, no.~6, June 2002.

\bibitem{mswfccm}
R.~C. de~Lamare, M.~Haardt, and R.~Sampaio-Neto, ``Blind adaptive constrained
  reduced-rank parameter estimation based on constant modulus design for cdma
  interference suppression,'' \emph{IEEE Transactions on Signal Processing},
  vol.~56, no.~6, June 2008.

\bibitem{spa}
R.~C. de~Lamare and R.~Sampaio-Neto, ``Minimum mean-squared error iterative
  successive parallel arbitrated decision feedback detectors for ds-cdma
  systems,'' \emph{IEEE Transactions on Communications}, vol.~56, no.~5, pp.
  778--789, 2008.

\bibitem{mfsic}
P.~Li, R.~C. De~Lamare, and R.~Fa, ``Multiple feedback successive interference
  cancellation detection for multiuser mimo systems,'' \emph{IEEE Transactions
  on Wireless Communications}, vol.~10, no.~8, pp. 2434--2439, 2011.

\bibitem{mbdf}
R.~de~Lamare, ``Adaptive and iterative multi-branch mmse decision feedback
  detection algorithms for multi-antenna systems,'' \emph{IEEE Transactions on
  Wireless Communications}, vol.~12, no.~10, pp. 5294--5308, 2013.

\bibitem{Krikidis}
I.~Krikidis, T.~Charalambous, and J.~Thompson, ``Buffer-aided relay selection
  for cooperative diversity systems without delay constraints,'' \emph{IEEE
  Transactions on Wireless Communications}, vol.~11, no.~5, pp. 1957--1967, May
  2012.

\end{thebibliography}

\end{document}